\documentclass[twocolumn,showpacs,preprintnumbers,amsmath,amssymb,prb]{revtex4-1}
\usepackage{graphicx}
\usepackage{dcolumn}
\usepackage{bm}
\usepackage{color}
\usepackage{physics}

\renewcommand{\r}{{\bm{r}}}

\def\lsim{\lower.35em\hbox{$\stackrel{\textstyle<}{\textstyle\sim}$}}
\def\gsim{\lower.35em\hbox{$\stackrel{\textstyle>}{\textstyle\sim}$}}

\begin{document}

\title{Magnetic phases from competing Hubbard and extended Coulomb interactions in twisted bilayer graphene}

\author{J. Gonz\'alez$^{1}$ and T. Stauber$^{2}$}

\affiliation{
$^{1}$ Instituto de Estructura de la Materia, CSIC, E-28006 Madrid, Spain\\
$^{2}$ Materials Science Factory,
Instituto de Ciencia de Materiales de Madrid, CSIC, E-28049 Madrid, Spain}
\date{\today}

\begin{abstract}
We implement a self-consistent Hartree-Fock approximation based on a microscopic model in real space, which allows us to consider the interplay between the Hubbard and the extended Coulomb interaction in twisted bilayer graphene at the magic angle. These two interactions tend to favor different symmetry breaking patterns, having therefore complementary roles in the regimes where one or the other dominates. We show that, for sufficiently large values of the on-site Hubbard repulsion, magic angle graphene has an antiferromagnetic ground state at the charge neutrality point, while at half-filling of the lowest valence band the state becomes fully spin-polarized. In general, a suitable screening of the extended Coulomb interaction is required to observe the magnetic state in either case, as otherwise the instabilities take place in the charge sector, preferentially in the form of time-reversal, chiral or valley symmetry breaking.
\end{abstract}
 
\maketitle

{\it Introduction.---} 
The discovery of superconductivity and correlated phases in twisted bilayer graphene (TBG) at the so-called magic angle\cite{Cao18b,Cao18a} has opened a new era in the investigation of strong electron correlations in layered materials
\cite{Yankowitz19,Codecido19,Shen19,Lu19,Chen19,Xu18,Volovik18,Yuan18,Po18,Roy18,Guo18,Dodaro18,Baskaran18,Liu18,Slagle18,Peltonen18,Kennes18,Koshino18,Kang18,Isobe18,Wu18b,Zhang18,Gonzalez19,Ochi18,Thomson18,Carr18,Guinea18,Zou18,Kang18,Kang19,Gonzalez19b}. There is indeed evidence that those phenomena may arise as a consequence of the strong electron-electron interaction in the bilayer, drawing a possible connection with the unconventional behavior of the copper-oxide superconductors.\cite{Park21,Hao21}

Some of the most prominent effects in TBG have to do with the opening of a gap in the electronic spectrum at integer fillings of the lowest valence and conduction bands\cite{Cao18b,Cao18a,Lu19}. It has been remarkable the observation of ferromagnetism at three-quarter filling of the lowest conduction band, with the concomitant breakdown of the spin and valley symmetries of the bilayer\cite{Sharpe19}. The gap seen at the charge neutrality point is also likely to arise from a dynamical breakdown of symmetry, by which the strong electronic interaction would destabilize the Dirac nodes in the spectrum.

There have been studies showing the feasibility that the electron interactions may induce different symmetry breaking patterns in TBG at the magic angle, although these have been mainly limited to the charge sector\cite{Xie19,Choi19,Kang19,Cea19,Rademaker19,Xie20,Liu21a,Liu21b,Zhang20,Gonzalez20,Kang20,Cea20,Lin20,Soejima20,Vafek20}. Magnetic instabilities have been less analyzed, perhaps due to the need to rely on a refined microscopic model discerning spin-dependent versus spin-independent interactions. In this regard, the relative strength of the spin-dependent Hubbard interaction (as compared to that of the extended Coulomb interaction) is the key element which governs the possibility of having a spin instability at different filling factors in TBG.    
   
In this paper, we implement a self-consistent Hartree-Fock approximation based on a microscopic model in real space, which allows us to consider the interplay between the Hubbard and the extended Coulomb interaction in TBG. These two interactions tend to favor different symmetry breaking patterns, having therefore complementary roles in the regimes where one or the other dominates. We find that the on-site Hubbard interaction governs in general the development of magnetic instabilities. We show in particular that, for sufficiently large values of the on-site Hubbard repulsion and close to the magic angle, TBG has an antiferromagnetic ground state at the charge neutrality point, while at half-filling of the lowest valence band the state becomes fully spin-polarized as predicted by a general theorem on the flat-band Hubbard model.\cite{Mielke91,MielkeTasaki93,Mielke93,Pons20} In general, a suitable screening of the extended Coulomb interaction is required to observe the magnetic state in either case, as otherwise the instabilities take place in the charge sector ---preferentially in the form of time-reversal, chiral or valley symmetry breaking.

{\it Hartree-Fock approximation.---}
We are going to focus our discussion on a twisted bilayer belonging to the set of commensurate superlattices with twist angle $\theta_i= \arccos ((3i^2+3i+0.5)/(3i^2+3i+1))$ \cite{Lopes07,Mele10}, taking in particular the representative with $i=28$ (twist angle $\theta \approx 1.16^\circ$). The Hamiltonian $H$ can be written as the sum of a non-interacting piece $H_0$ and the term $H_{\rm int}$ containing the Hubbard and Coulomb interactions
\begin{align}
H = H_0 + H_{\rm int}\;.
\end{align}
We represent $H_0$ in the form of a tight-binding Hamiltonian, adopting the parametrization already used in Ref. \onlinecite{Gonzalez20}. In the moir\'e superlattice with electrons sitting at lattice sites $\r_i$ with spin $\sigma = \uparrow, \downarrow$, the matrix representation $( H_0 )_{i\sigma,j\sigma}$ can be exactly diagonalized, leading to eigenvalues $\varepsilon_{a\sigma}^0$ and eigenvectors $\phi_{a\sigma}^0 (\r_i)$. In the zero-frequency (static) limit, the non-interacting electron propagator $G_0 $ is just the inverse of $H_0$, and it can be written as
\begin{align}
\left(  G_0  \right)_{i\sigma,j\sigma} = -\sum_a \frac{1}{\varepsilon_{a\sigma}^0}  \phi_{a\sigma}^0 (\r_i)  \phi_{a\sigma}^0 (\r_j)^*\;.
\end{align}

The Hartree-Fock approximation relies on the assumption that the full electron propagator $G$ can be represented in terms of a modified set of eigenvalues $\varepsilon_{a\sigma}$ and eigenvectors $\phi_{a\sigma} (\r_i)$, in such a way that in the static limit
\begin{align}
\left(  G  \right)_{i\sigma,j\sigma} = -\sum_a \frac{1}{\varepsilon_{a\sigma}}  \phi_{a\sigma} (\r_i)  \phi_{a\sigma} (\r_j)^*\;.
\label{hf}
\end{align}
Furthermore, the relation between $G$ and $G_0$ is given by the electron self-energy $\Sigma $ according to the Dyson equation
\begin{align}
G^{-1} = G_0^{-1} - \Sigma\;.
\label{dyson}
\end{align}  
In the Hartree-Fock approximation, the many-body diagrammatics implies that $\Sigma $ can be expressed in terms of the set of $\varepsilon_{a\sigma}$ and $\phi_{a\sigma} (\r_i)$. It turns out that, in the static limit, 
\begin{align}
\left( \Sigma  \right)_{i\sigma,j\sigma}  = &  \; 2 \mathbb{I}_{ij} \:  \sideset{}{'}\sum_a  \sum_{l, \sigma' } v_{\sigma \sigma'} (\r_i-\r_l)   \left|\phi_{a\sigma'} (\r_l)\right|^2      \notag    \\ 
    &  - v_{\sigma \sigma} (\r_i-\r_j)  \sideset{}{'}\sum_a \phi_{a\sigma} (\r_i)  \phi_{a\sigma} (\r_j)^*\;,
\label{selfe}
\end{align}
where $v_{\sigma \sigma'} (\r)$ is the interaction potential between electron densities with spin $\sigma $ and $\sigma' $ and the prime means that the sum is to be carried over the occupied levels \cite{Fetter71}.

The problem of finding the set of $\varepsilon_{a\sigma}$ and $\phi_{a\sigma} (\r_i)$ amounts then to solving the self-consistent equation given by (\ref{dyson}) and (\ref{selfe}). This can be achieved in practice by means of a recursive procedure, in which the self-energy is built at each step from approximate eigenvalues and eigenvectors obtained in the previous iteration.  

One of the advantages of applying the Hartree-Fock approximation in real space is the possibility to discern the contribution of different interactions to the potential $v_{\sigma \sigma'} (\r)$. We have for instance a term $H_{\rm C}$ in the interaction Hamiltonian corresponding to the extended Coulomb interaction, which we will take as suitably screened by nearby metallic gates in order to make contact with typical experimental setups. Thus, we can express in terms of electron creation (annihilation) operators $a_{i\sigma}^+$ ($a_{i\sigma}$)
\begin{align}  
H_{\rm C} = \frac{1}{2} \sum_{i,j,\sigma,\sigma'} a_{i\sigma}^{\dagger}a_{i\sigma} \: v_{\rm C} (\r_i-\r_j) \: a_{j\sigma'}^{\dagger}a_{j\sigma'}\;,
\end{align}
where we take the potential appropriate to the case of top and bottom metallic gates\cite{Throckmorton12}, each at a distance $d = \xi /2$ from the twisted bilayer,
\begin{align}
v_{\rm C} (\r )      =    \frac{e^2}{4\pi \epsilon}   \frac{2\sqrt{2} \: e^{-\pi r/\xi }}{\xi \sqrt{r/\xi }}\;.
\label{vscr}
\end{align}
Moreover, we also take into account the interaction coming from the on-site repulsion of electrons sitting at the same carbon atom. This contributes to $H_{\rm int}$ with the Hubbard term
\begin{align}  
H_{\rm U} =  U \: \sum_{i} a_{i\uparrow}^{\dagger}a_{i\uparrow}  \: a_{i\downarrow}^{\dagger}a_{i\downarrow}\;.
\end{align}

The term $H_{\rm U}$ can be viewed as a prescription to define the Coulomb interaction in the limit $\r \rightarrow 0$, which cannot be obtained from (\ref{vscr}). The on-site repulsion $U$ is actually a very relevant parameter in the subsequent discussion since the Hubbard term is the spin-dependent part of the interaction. The prevalence of the magnetic phases turns out to be dictated then by the value of $U$, as we see in what follows.

{\it Symmetry breaking at the charge neutrality point.---}
The most distinctive experimental feature observed at the charge neutrality point of TBG is the opening of a gap in the electronic spectrum. This can be attributed to the effect of dynamical symmetry breaking which, for sufficiently strong electron-electron interaction, destabilizes the Dirac nodes at the $K$ points of the moir\'e Brillouin zone. Such an effect proceeds typically through the development of a staggered density in the charge or the spin sector. The different symmetry breaking patterns can be built from the matrix elements 
\begin{align}
h_{ij}^{(\sigma )} =  \sideset{}{'}\sum_a \phi_{a\sigma} (\r_i) \phi_{a\sigma} (\r_j)^*\;,
\end{align}
where the prime means again that the sum is only over occupied states. In TBG, we have sublattices $A_1, B_1$ for the top carbon layer and $A_2, B_2$ for the bottom layer. Thus, we have the order parameters
\begin{align}
C_{\pm\sigma} = & \sum_{i \in A_1} h_{ii}^{(\sigma )}  - \sum_{i \in B_1} h_{ii}^{(\sigma )} 
      \pm  \left( \sum_{i \in A_2} h_{ii}^{(\sigma )} - \sum_{i \in B_2} h_{ii}^{(\sigma )}   \right) \;.
\end{align}
The condensation of a staggered spin density (signaling antiferromagnetic order) may be characterized by a nonvanishing value of $C_{+\uparrow} - C_{+\downarrow}$, while the prevalence of a nonzero value of $C_{+\uparrow} + C_{+\downarrow}$ is instead the signature of staggered charge order, with the consequent chiral symmetry breaking.

As already pointed out, the Hubbard term is the only source of spin-dependent interaction in our model, so that the balance between the charge and the spin order is governed by the relative strength of the on-site repulsion $U$. This quantity is assumed to have a value of the order of $\sim 8$ eV in graphene, but in TBG it may be significantly reduced due to internal screening from the narrow bands. The value of the renormalized parameter depends sensitively on the degree of proximity to the magic angle, and here we are going to consider two different instances corresponding to moderate internal screening (with an effective value of $U = 4$ eV) and strong internal screening (effective value of $U = 0.5$ eV).

\begin{figure}[t]
(A)$ $\hspace{8cm}$ $\\
\includegraphics[width=0.8\columnwidth]{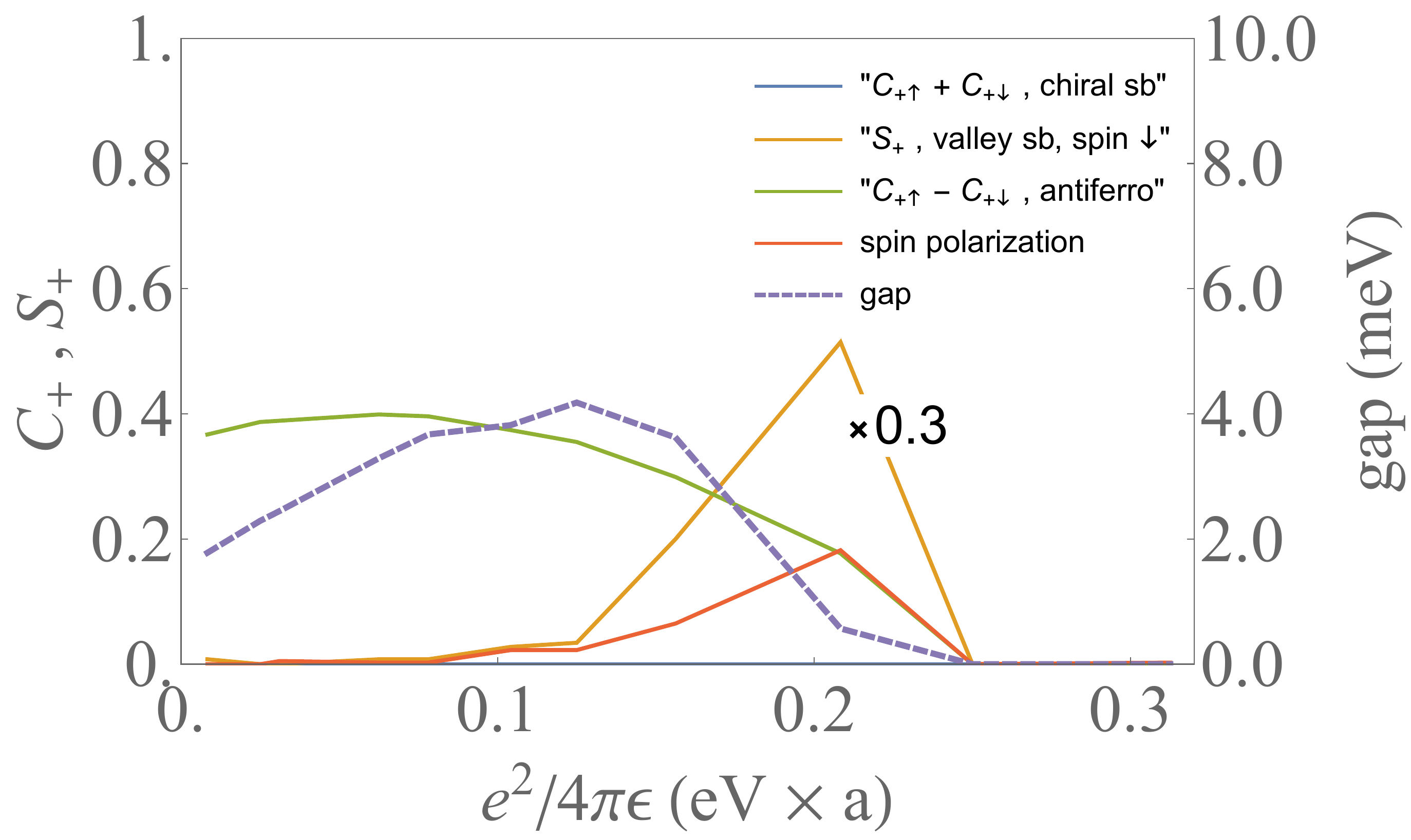}\\
(B)$ $\hspace{8cm}$ $\\
\includegraphics[width=0.8\columnwidth]{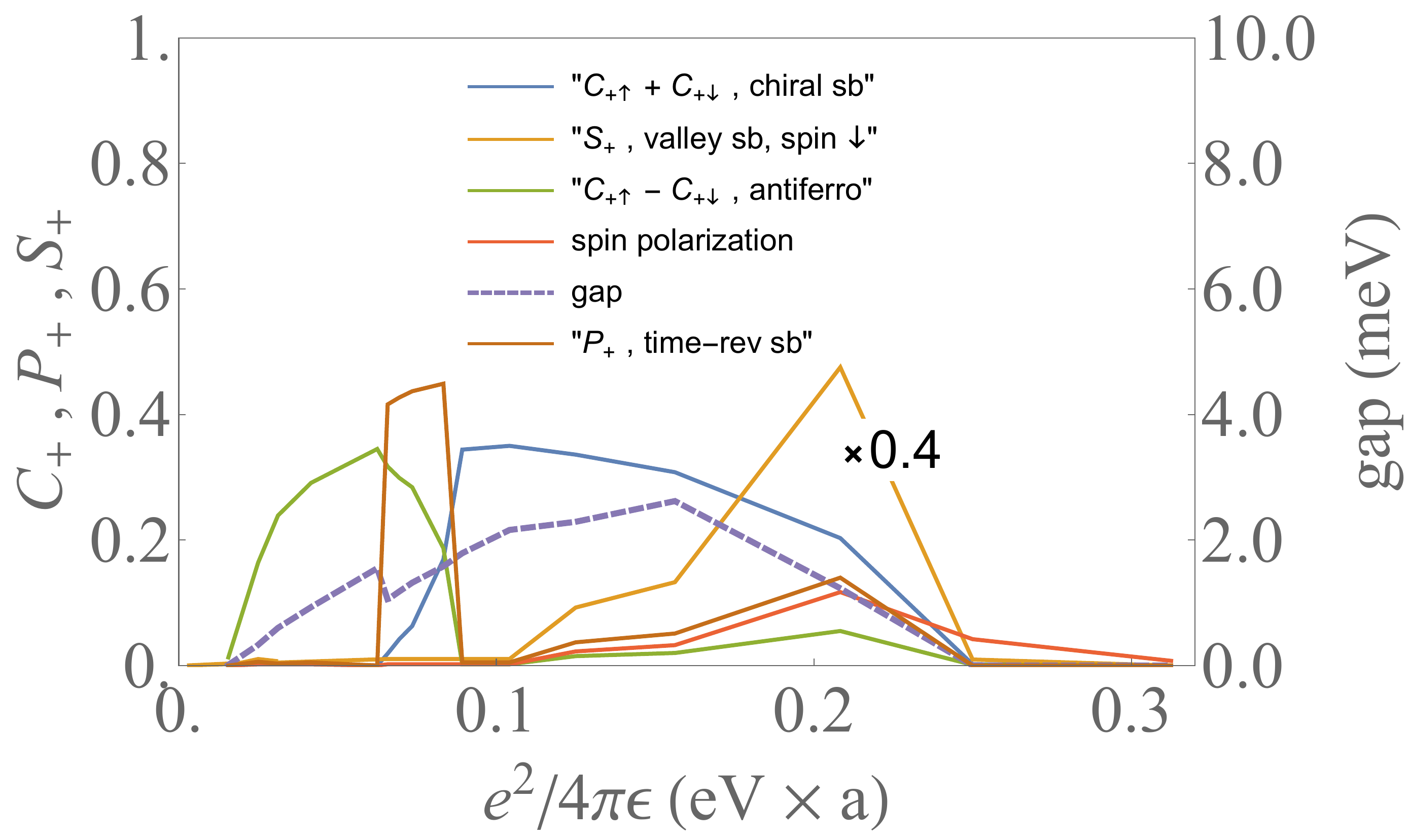}\\
\caption{Phase diagrams showing the order parameters of the dominant symmetry breaking patterns at the charge neutrality point of twisted bilayer graphene with $i=28$ (twist angle $\theta \approx 1.16^\circ$) for two different values of the on-site Hubbard coupling $U = 4.0$ eV (A) and 0.5 eV (B). The $x$-axis corresponds to the coupling of the Coulomb potential (in units where $a$ is the C-C distance).}
\label{one}
\end{figure}

For the effective value $U = 4$ eV, we find that the development of a staggered spin density (antiferromagnetic order) prevails over chiral symmetry breaking, independent of the strength of the Coulomb interaction. This is shown in the phase diagram of Fig. \ref{one}(A), where the $x$-axis represents the strength of the potential (\ref{vscr}), for a setup with $\xi = 10$ nm. We observe that the antiferromagnetic phase is preserved down the limit of a pure Hubbard interaction ($\epsilon \rightarrow \infty $).

When the antiferromagnetic signal decreases to the right of the diagram in Fig. \ref{one}(A), we find however the onset a different order parameter corresponding to valley symmetry breaking in TBG. This may be characterized by the circulation of the matrix elements $h_{ij}$ along the three nearest neighbors $i_1, i_2$ and $i_3$ of each atom $i$, with clockwise orientation. We may have a nonvanishing flux inside each loop indicating the breakdown of time-reversal invariance, but with opposite sign in the two sublattices $A$ and $B$. This condensation is measured by the order parameters \begin{align}
S_{\pm\sigma} &= {\rm Im} \left( \sum_{i \in A_1}  \left(   h_{i_1 i_2}^{(\sigma )} h_{i_2 i_3}^{(\sigma )} h_{i_3 i_1}^{(\sigma )}   \right)^{\frac{1}{3}}
          - \sum_{i \in B_1}  \left(   h_{i_1 i_2}^{(\sigma )} h_{i_2 i_3}^{(\sigma )} h_{i_3 i_1}^{(\sigma )}   \right)^{\frac{1}{3}}   \right.    \notag   \\
     &  \left. \pm    \sum_{i \in A_2}  \left(   h_{i_1 i_2}^{(\sigma )} h_{i_2 i_3}^{(\sigma )} h_{i_3 i_1}^{(\sigma )}   \right)^{\frac{1}{3}}
          \mp \sum_{i \in B_2} \left(    h_{i_1 i_2}^{(\sigma )} h_{i_2 i_3}^{(\sigma )} h_{i_3 i_1}^{(\sigma )}  \right)^{\frac{1}{3}}  \right)\;.
          \label{haldS}
\end{align}
The phase realized to the right of the diagram in Fig. \ref{one}(A) corresponds to nonvanishing $S_{+\sigma}$ for the less populated spin $\sigma $. In the continuum theory of Dirac fermions, this breakdown of symmetry translates into the generation of a term proportional to the identity in pseudospin space. This does not open a gap in the Dirac cones at the $K$ point, but instead it leads to a different shift in the energy of the cones in the two valleys of the twisted bilayer, with the consequent valley symmetry breaking. 

On the other hand, if we take an effective value of $U = 0.5$ eV, the phase diagram shows a different competition between symmetry breaking patterns as seen in Fig. \ref{one}(B). For that value of $U$, we observe that there is no magnetic instability in the limit of a pure Hubbard interaction $\epsilon \rightarrow \infty $. However, there is still an antiferromagnetic phase in the weak-coupling regime of the extended Coulomb interaction, where this is presumably reinforcing the Hubbard interaction to produce the staggered spin order. 

For larger Coulomb interaction, we find next a mixed phase where there is chiral symmetry breaking (staggered charge order) for one of the spin polarizations, while for the other polarization, the state corresponds to a Chern insulator with the order parameter   
\begin{align}
P_{+\sigma} &= {\rm Im} \left( \sum_{i \in A_1}  \left(   h_{i_1 i_2}^{(\sigma )} h_{i_2 i_3}^{(\sigma )} h_{i_3 i_1}^{(\sigma )}   \right)^{\frac{1}{3}}
          + \sum_{i \in B_1}  \left(   h_{i_1 i_2}^{(\sigma )} h_{i_2 i_3}^{(\sigma )} h_{i_3 i_1}^{(\sigma )}   \right)^{\frac{1}{3}}   \right.    \notag   \\
     &  \left. +    \sum_{i \in A_2}  \left(   h_{i_1 i_2}^{(\sigma )} h_{i_2 i_3}^{(\sigma )} h_{i_3 i_1}^{(\sigma )}   \right)^{\frac{1}{3}}
          + \sum_{i \in B_2} \left(    h_{i_1 i_2}^{(\sigma )} h_{i_2 i_3}^{(\sigma )} h_{i_3 i_1}^{(\sigma )}  \right)^{\frac{1}{3}}  \right)\;.
\label{hald}
\end{align}

For a yet stronger Coulomb interaction, we turn into a phase with chiral symmetry breaking in the two spin polarizations. This phase as well as the preceding Chern insulating phase are already present in the phase diagram obtained for spin-independent interactions in TBG\cite{Gonzalez20}. We may interpret therefore that those two phases reflect the regime of prevalence of the extended Coulomb interaction. In any case, we observe that there is still a residual effect of the Hubbard interaction to the right of the phase diagram in Fig. \ref{one}(B), leading to a slight spin polarization and a concomitant phase with valley symmetry breaking.

Finally, for the smallest values of the dielectric constant $\epsilon $, there is no signal of symmetry breaking in the phase diagram of Fig. \ref{one}(B). Nevertheless, this is a consequence of the fact that, for such a strong-coupling regime, the Fermi level departs at charge neutrality from the neighborhood of the Dirac nodes at the $K$ points of the moir\'e Brillouin zone. This explains the apparent discrepancy with the phase diagrams in Ref. \onlinecite{Gonzalez20}, where the phases have been always found by constraining the Fermi level to fall between the upper and lower Dirac cones.

{\it Symmetry breaking at half-filling of the lowest valence band.---}
As the magic angle is approached and the low-energy bands flatten, the one-particle density of states becomes progressively higher in the lowest valence and conduction bands of TBG. The question of symmetry breaking becomes then relevant and, in particular, whether a spin-polarized state may arise due to interaction effects. In this regard, the key role is played by the effective value after screening of the Hubbard coupling $U$. As in the previous section, we are going to consider again two different instances, corresponding to moderate and strong internal screening of the interaction and with respective effective values of the Hubbard coupling $U = 4$ eV and 0.5 eV.

\begin{figure}[t]
(A)$ $\hspace{8cm}$ $\\
\includegraphics[width=0.8\columnwidth]{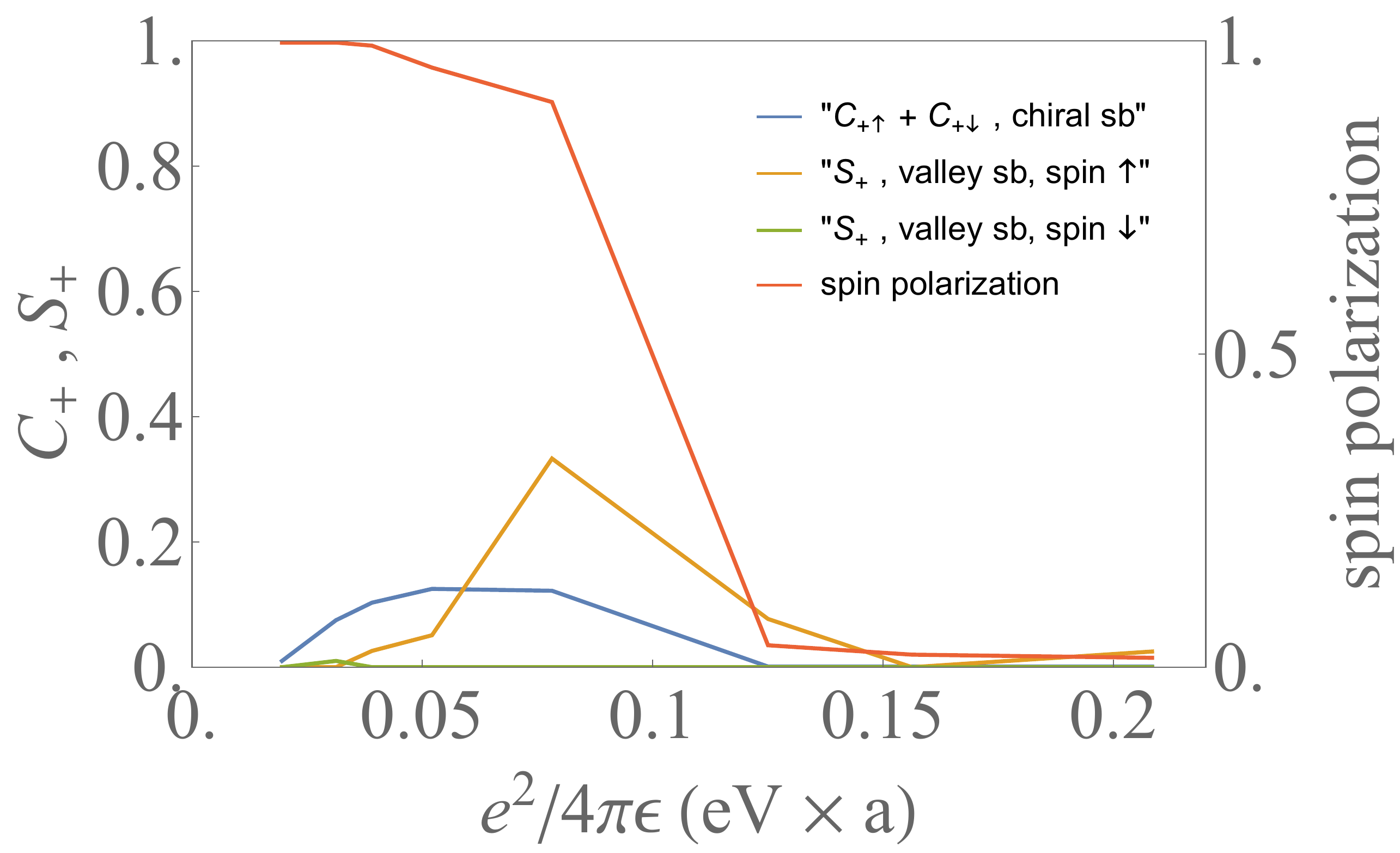}\\
(B)$ $\hspace{8cm}$ $\\
\includegraphics[width=0.8\columnwidth]{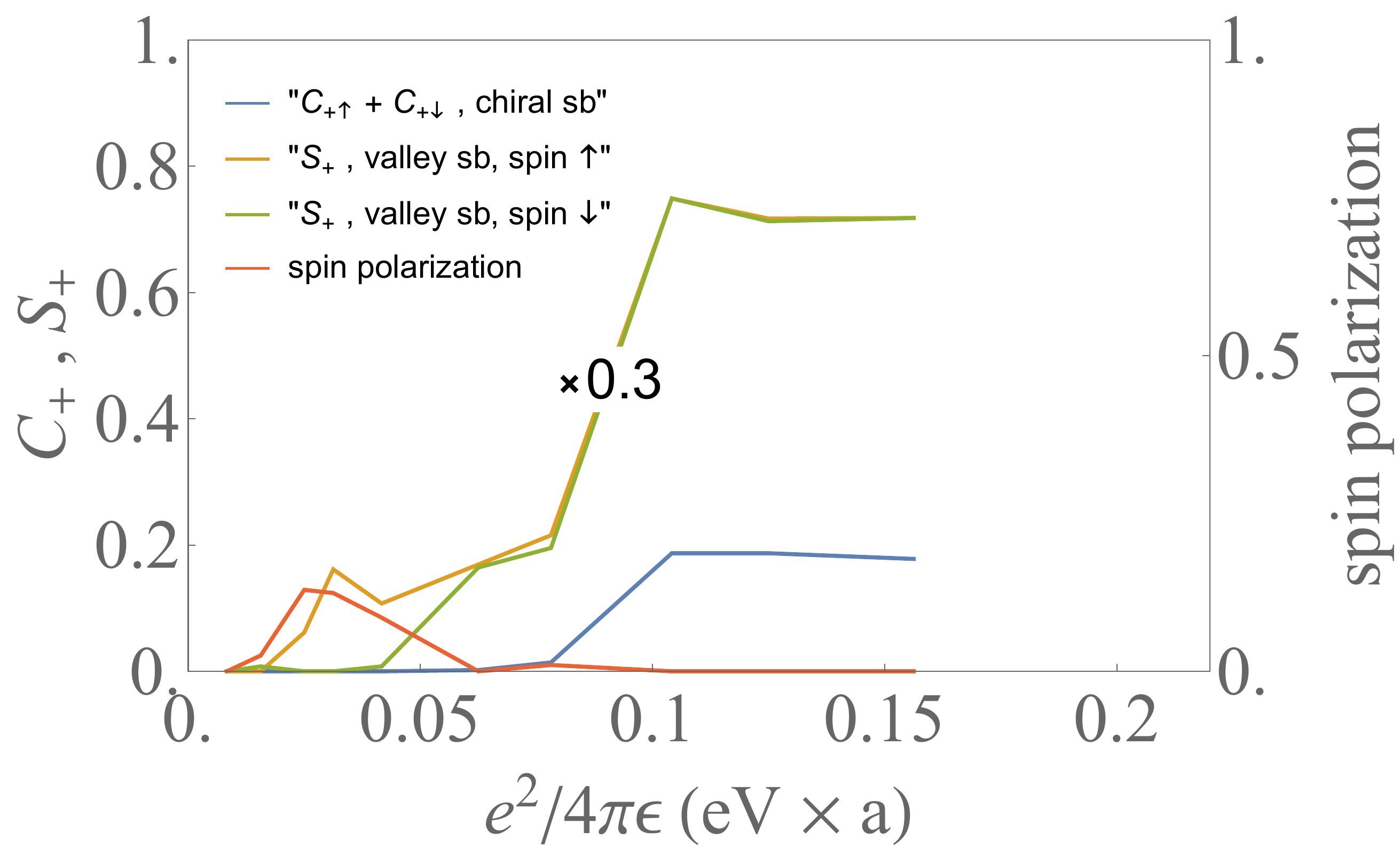}\\
\caption{Phase diagrams showing the order parameters of the dominant symmetry breaking patterns at half-filling of the lowest valence band of twisted bilayer graphene with $i=28$ (twist angle $\theta \approx 1.16^\circ$) for two different values of the on-site Hubbard coupling $U = 4.0$ eV (A) and 0.5 eV (B). The $x$-axis corresponds to the coupling of the Coulomb potential (in units where $a$ is the C-C distance).}
\label{two}
\end{figure}

\begin{figure}[t]
(A)$ $\hspace{8cm}$ $\\
\includegraphics[width=0.99\columnwidth]{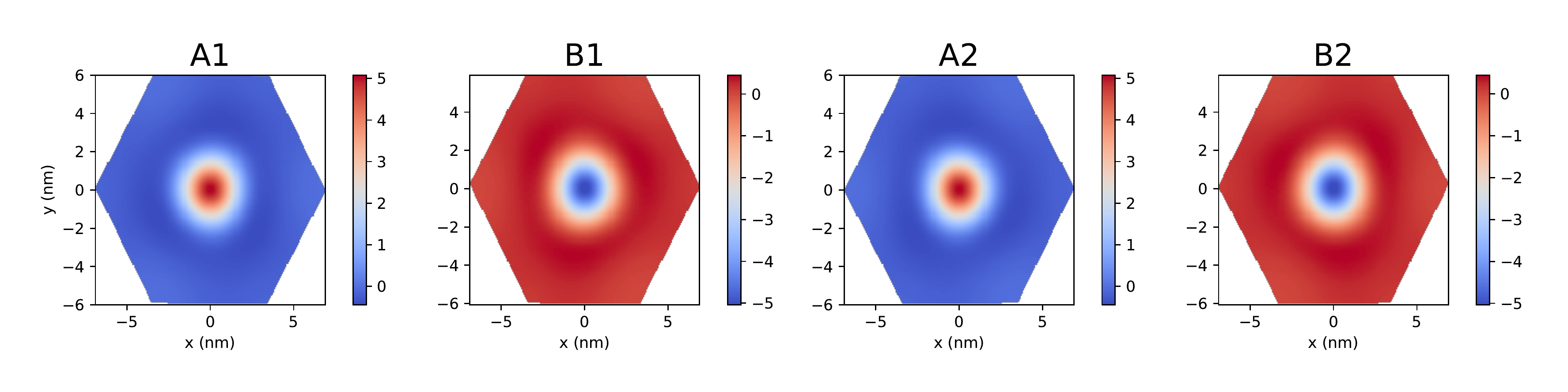}\\
(B)$ $\hspace{8cm}$ $\\
\includegraphics[width=0.99\columnwidth]{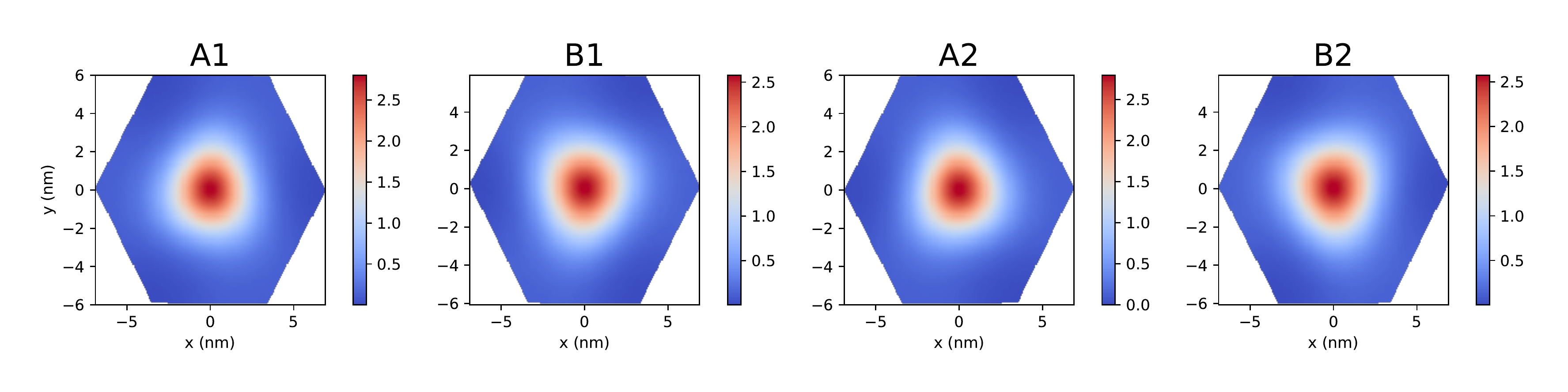}\\
\caption{Spin density at the neutrality point (A) and at half-filling of the lowest valence band (B) of twisted bilayer graphene with $i=28$ (twist angle $\theta \approx 1.16^\circ$) for an on-site Hubbard coupling $U = 4.0$ eV. We show the spin densities separately for the $X$-lattice sites of layer $n$ ($Xn$)  with $X=A,B$ and $n=1,2$, in units of the density $1/A_i$ of one electron inside the unit cell of the moir\'e lattice with $A_i=(3i^3+3i+1)\bar{a}^2$ and $\bar{a}=0.246$ nm.}
\label{three}
\end{figure}

For an effective on-site repulsion $U = 4$ eV, the Hubbard interaction is strong enough to produce a full spin polarization in the ground state of the system. This is shown in the phase diagram of Fig. \ref{two}(A), which represents the different phases for growing strength of the extended Coulomb interaction. The full spin polarization is localized around the AA-stacked region of the unit cell, see Fig. \ref{three}, and takes place along with the complete splitting of the low-energy bands for spin up and spin down. This means that, at half-filling, the polarized electrons populate the states up to the level of the Dirac nodes of the filled band. These are then susceptible of being destabilized by the electronic interaction, as it actually happens with the development of chiral symmetry breaking seen in the phase diagram of Fig. \ref{two}(A). The diagram also shows a clear competition between the effects of the Hubbard and the Coulomb interaction, which leads to the suppression of the spin polarization for sufficiently small values of the dielectric constant $\epsilon $.

The same effect of full spin polarization is found for $U = 2$ eV in the limit $\epsilon \rightarrow \infty $, but for $U = 0.5$ eV the magnetic ground state is lost however, as shown in Fig. \ref{two}(B). For that value of $U$, we also observe that the main signature of symmetry breaking corresponds to the prominent growth of the order parameter $S_+$ defined in Eq. (\ref{haldS}). Therefore, the main physical effect is that of valley symmetry breaking, which is driven by the extended Coulomb interaction and grows large for small values of the dielectric constant.

{\it Conclusion.---}
In this paper we have reported self-consistent Hartree-Fock calculations for magic angle twisted bilayer graphene including both on-site Hubbard and extended Coulomb interactions. We have found that, for realistic parameters, the dynamical generation of a gap is mainly due to the extended Coulomb interaction. The system is thus far away from the well-known semimetal-antiferromagnetic Mott insulator transition. In this context, let us remark that we have carried out the calculations without breaking the three-fold rotational symmetry of the twisted bilayer. Nevertheless, the presented conclusions are still relevant because a nematic instability does not usually open a gap at the Dirac cones. This means that some of the symmetry breaking patterns discussed in the paper must be at work to produce the opening of the gap observed at the charge neutrality point.

We have been mainly interested in magnetic phases, and we have found a crossover from antiferromagnetic to ferromagnetic order when doping the system from the neutrality point to the half-filled first valence band. The resulting spin-densities are localized around the AA-stacked regions and for one sub-lattice and layer they can be as large as 5 times the density of one electron per moir\'e unit cell. We thus believe that this ^^ ^^ magnetic lattice" should be detectable in future experiments, providing a way of testing the balance between Hubbard and extended Coulomb interactions in the twisted bilayer.

{\it Acknowledgments.}
This work has been supported by Spain's MINECO under Grant No. FIS2017-82260-P as well as by the CSIC Research Platform on Quantum Technologies PTI-001. The access to computational resources of CESGA (Centro de Supercomputaci\'on de Galicia) is also gratefully acknowledged.

\bibliography{hfspin4.bib}

\end{document}